\documentclass[aps,prc,superscriptaddress,showpacs,showkeys,twocolumn]{revtex4-1}

\usepackage{graphicx}
\usepackage{dcolumn}
\usepackage{bm}
\usepackage{hyperref}
\usepackage{color}
\usepackage{longtable} 
\usepackage{multirow}
\usepackage{amsmath}
\usepackage[T1]{fontenc}
\usepackage{textcomp}

\begin{document}


\title{Measurement of Lifetimes in \texorpdfstring{$^{23}$Mg}{23Mg}}

\author{O.\ S.\ Kirsebom}
\email[]{oliskir@phys.au.dk}
\affiliation{Department of Physics and Astronomy, Aarhus University, DK-8000 Aarhus C, Denmark}


\author{P.\ Bender}
\affiliation{Science Division, TRIUMF, Vancouver, British Columbia, V6T 2A3, Canada}

\author{A.\ Cheeseman}
\affiliation{Science Division, TRIUMF, Vancouver, British Columbia, V6T 2A3, Canada}

\author{G.\ Christian}
\affiliation{Science Division, TRIUMF, Vancouver, British Columbia, V6T 2A3, Canada}

\author{R.\ Churchman}
\affiliation{Science Division, TRIUMF, Vancouver, British Columbia, V6T 2A3, Canada}

\author{D.\ S.\ Cross}
\affiliation{Department of Chemistry, Simon Fraser University, Burnaby, British Columbia, V5A 1S6, Canada}

\author{B.\ Davids}
\affiliation{Science Division, TRIUMF, Vancouver, British Columbia, V6T 2A3, Canada}

\author{L.\ J.\ Evitts}
\affiliation{Science Division, TRIUMF, Vancouver, British Columbia, V6T 2A3, Canada}
\affiliation{Department of Physics, University of Surrey, Guildford GU2 5XH, United Kingdom}

\author{J.\ Fallis}
\affiliation{Science Division, TRIUMF, Vancouver, British Columbia, V6T 2A3, Canada}

\author{N.\ Galinski}
\affiliation{Science Division, TRIUMF, Vancouver, British Columbia, V6T 2A3, Canada}
\affiliation{Department of Physics, Simon Fraser University, Burnaby, British Columbia, V5A 1S6, Canada}

\author{A.\ B.\ Garnsworthy}
\affiliation{Science Division, TRIUMF, Vancouver, British Columbia, V6T 2A3, Canada}

\author{G.\ Hackman}
\affiliation{Science Division, TRIUMF, Vancouver, British Columbia, V6T 2A3, Canada}

\author{J.\ Lighthall}
\affiliation{Science Division, TRIUMF, Vancouver, British Columbia, V6T 2A3, Canada}

\author{S.\ Ketelhut}
\affiliation{Science Division, TRIUMF, Vancouver, British Columbia, V6T 2A3, Canada}

\author{P.\ Machule}
\affiliation{Science Division, TRIUMF, Vancouver, British Columbia, V6T 2A3, Canada}

\author{D.\ Miller}
\affiliation{Science Division, TRIUMF, Vancouver, British Columbia, V6T 2A3, Canada}

\author{C.\ R.\ Nobs}
\altaffiliation{Present address: School of Computing, Engineering and Mathematics, University of Brighton, Brighton, BN2 4GJ, United Kingdom.} 
\affiliation{Science Division, TRIUMF, Vancouver, British Columbia, V6T 2A3, Canada}
\affiliation{Department of Physics, University of Surrey, Guildford GU2 5XH, United Kingdom}

\author{C.\ J.\ Pearson}
\affiliation{Science Division, TRIUMF, Vancouver, British Columbia, V6T 2A3, Canada}

\author{M.\ M.\ Rajabali}
\affiliation{Science Division, TRIUMF, Vancouver, British Columbia, V6T 2A3, Canada}

\author{A.\ J.\ Radich}
\affiliation{Science Division, TRIUMF, Vancouver, British Columbia, V6T 2A3, Canada}
\affiliation{Department of Physics, University of Guelph, Guelph, Ontario, N1G 2W1, Canada}

\author{A.\ Rojas}
\altaffiliation{Present address: Occupational Health and Safety Division, Government of Saskatchewan, Regina, Saskatchewan, S4P 4W1, Canada.}
\affiliation{Science Division, TRIUMF, Vancouver, British Columbia, V6T 2A3, Canada}

\author{C.\ Ruiz}
\affiliation{Science Division, TRIUMF, Vancouver, British Columbia, V6T 2A3, Canada}

\author{A.\ Sanetullaev}
\affiliation{Science Division, TRIUMF, Vancouver, British Columbia, V6T 2A3, Canada}
\affiliation{Astronomy and Physics Department, Saint Mary's University, Halifax, Nova Scotia, B3H 3C3, Canada}

\author{C.\ D.\ Unsworth}
\affiliation{Science Division, TRIUMF, Vancouver, British Columbia, V6T 2A3, Canada}

\author{C.\ Wrede}
\affiliation{Department of Physics and Astronomy, Michigan State University, East Lansing, Michigan 48824, USA}
\affiliation{National Superconducting Cyclotron Laboratory, Michigan State University, East Lansing, Michigan 48824, USA}


\date{\today}

\begin{abstract}
Several lifetimes in $^{23}$Mg have been determined for the first time using the Doppler-shift attenuation method. A Monte Carlo simulation code has been written to model the $\gamma$-ray line shape. An upper limit of $\tau < 12$~fs at the 95\% C.L.\ has been obtained for the astrophysically important 7787~keV state.
\end{abstract}


\maketitle


\section{Introduction}\label{sec:intro}%


The radionuclide $^{22}$Na is thought to be produced in significant amounts in ONe-type classical novae~\cite{jose06}. It decays by $\beta^+\nu_e$ emission to the first excited state in $^{22}$Ne with a mean lifetime of 3.75~years. The daughter nucleus quickly relaxes to the ground state by emitting a $\gamma$ ray with a characteristic energy of 1275~keV. The lifetime of 3.75 years is sufficiently short to ensure that the ejected $^{22}$Na is still close to the nova by the time the decay occurs, making it possible to correlate the $\gamma$-ray intensity with other observational properties, yet sufficiently long for the $^{22}$Na to survive beyond the opaque phase of the explosion, allowing the $\gamma$-ray to be observed. These fortunate circumstances make $^{22}$Na the most promising candidate for detection of nuclear $\gamma$-ray lines from novae~\cite{clayton1974}. Nova models predict $^{22}$Na $\gamma$-ray fluxes 1 order of magnitude below the observational upper limit~\cite{iyudin95}. While detection does not appear to be imminent, it seems reasonable to assume that it will happen within the foreseeable future~\cite{hernanz13}, and consequently reliable estimates of $^{22}$Na yields in novae are desirable. 
Nova production of $^{22}$Na may also be relevant for explaining the nonstandard abundance of $^{22}$Ne observed in Ne-E meteorites~\cite{black1972}.

Under the conditions found in novae $^{22}$Na is mainly destroyed by proton capture. Two direct measurements of the proton-capture cross section have been reported. Such measurements are very challenging because the target material is radioactive. The first measurement was performed at Ruhr-Universit\"at Bochum, Germany, in two attempts with results published in 1990~\cite{seuthe90} and 1996~\cite{stegmuller96}. The second measurement was performed at the University of Washington in Seattle with results published in 2010~\cite{sallaska10} and 2011~\cite{sallaska11}. Both studies conclude that at peak nova temperatures the reaction is dominated by a resonance at a proton energy of $E_p=213$~keV, corresponding to an excitation energy of $E_x=7787$~keV in the compound nucleus $^{23}$Mg. (At the very highest nova temperatures a resonance at 288 keV also makes a significant contribution.) However, the resonance strengths reported for the 213 keV resonance differ substantially: 
\begin{align*}
\omega\gamma &= 1.8\pm 0.7\textrm{ meV} \quad \textrm{(Bochum)} \\
\omega\gamma &= 5.7^{+1.6}_{-0.9}\textrm{ meV} \quad \textrm{(Seattle)} 
\end{align*}
This leads to an uncertainty of about a factor of 2 in the amount of $^{22}$Na ejected from novae~\cite{sallaska11, parikh2014}.

Since $p+{}^{22}$Na and $\gamma+{}^{23}$Mg are the only open channels, the resonance strength is given by
\begin{equation}\label{eq : strength}
\omega \gamma = \frac{2J+1}{(2J_p+1)(2J_{\textrm{Na}}+1)} \frac{\hbar}{\tau} B_p(1-B_p)
\end{equation}
where $J_p=1/2$ and $J_{\textrm{Na}}=3$ are the spins of the proton and $^{22}$Na, respectively, $J$ is the spin of the resonance, $\tau$ is the mean lifetime, and $B_p$ is the branching ratio for proton decay. Thus a measurement of $J$, $\tau$, and $B_p$ provides an indirect determination of the resonance strength.
The spin-parity and mean lifetime of the 7787~keV state have been determined to be $J^{\pi}=7/2^{(+)}$ and $\tau=10\pm 3$~fs, respectively, in a measurement of the fusion-evaporation reaction $^{12}$C($^{12}$C$,n){}^{23}$Mg~\cite{jenkins04, jenkins13}. 
The branching ratio for proton decay has been determined to be $B_p = 0.037\pm 0.07$ in a measurement of the $\beta p$ decay of $^{23}$Al~\cite{saastamoinen11}, which also provides compelling evidence for a $J^{\pi}=(7/2)^+$ assignment.
Thus the indirect approach gives
\begin{align*}
\omega \gamma &= 1.3\pm 0.5\textrm{ meV} \quad \textrm{(Indirect, $J=7/2$)}
\end{align*}
with the uncertainty obtained by adding the uncertainties on $\tau$ (30\%) and $B_p$ (19\%) in quadrature. This value agrees with the old and less precise result of the Bochum group, but disagrees with the more recent and more precise result of the Seattle group. 
To resolve the disagreement we set out to remeasure $\tau$. %

We note that recently a different spin-parity assignment for the 7787~keV state has been proposed by Tripathi {\it et al.}~\cite{tripathi2013}. Based on a comparison of the experimentally determined $\beta$-decay strength of the 7787~keV state to the $\beta$-decay strength predicted by a shell-model calculation, Tripathi {\it et al.}\ argue that $J^{\pi}=5/2^+$ is a more plausible assignment, which gives 
\begin{align*}
\omega \gamma &= 1.0\pm 0.4\textrm{ meV} \quad \textrm{(Indirect, $J=5/2$)}
\end{align*}
aggravating the discrepancy with respect to the Seattle measurement.

\begin{figure}
\includegraphics[width=0.95\linewidth, angle=0, clip=true, trim= 0 0 0 0]{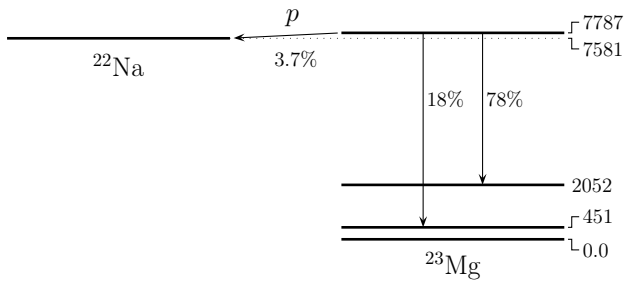}%
\caption{Decay scheme of the astrophysically important 7787~keV state. The levels are labeled by their excitation energy in keV.}
\label{fig:levelscheme}
\end{figure}

In a recent measurement of the $\beta \gamma$ spectrum of $^{23}$Al~\cite{iacob06, zhai_thesis} the 7787~keV state is observed to decay to the 451 keV and 2052 keV states with relative intensities of 81(4)\% and 19(4)\% and $\gamma$-ray energies of 7335.2(6)~keV and 5735.4(7)~keV. The excitation energy of the 7787~keV state is determined to be $E_x=7787.2(6)$~keV, which is the most precise value to date and the value that we adopt in the present study. %
While originally only published in a PhD thesis~\cite{zhai_thesis}, this value has since been quoted in a peer-reviewed publication~\cite{saastamoinen11}.
We further note that the value agrees with the value of $E_x=7785.7\pm 1.1$~keV obtained in the $^{12}$C($^{12}$C$,n){}^{23}$Mg experiment~\cite{jenkins04, jenkins13}, where only the $7787\,\textrm{keV}\rightarrow 451\,\textrm{keV}$ transition was observed.
(The value given in Ref.~\cite{jenkins04} is $E_x=7784.6\pm 1.1$~keV, but this value appears not to have been corrected for the nuclear recoil.)
A simplified level scheme is shown in Fig.~\ref{fig:levelscheme}.%

\section{Experimental Technique}\label{sec:technique}%
We employ the Doppler-shift attenuation method (DSAM)~\cite{branford73, alexander78}. The excited nucleus is produced at high speed and slows down in a dense medium. If the lifetime is sufficiently short, the nucleus will decay before it comes to rest, and hence the energy of the emitted $\gamma$-ray will be Doppler-shifted according to,
\begin{equation}
E_{\gamma} = E_{\gamma,0} \; \frac{ \left( \, 1-\beta^2 \, \right)^{1/2} }{ 1-\beta \cos \theta_{\gamma} } \; , 
\end{equation}
where $E_{\gamma,0}$ is the $\gamma$-ray energy at rest, $\beta$ is the instantaneous speed of the nucleus in units of $c$, and $\theta_{\gamma}$ is the angle of emission of the $\gamma$ ray relative to the trajectory of the nucleus. A shorter lifetime leaves less time for the nucleus to slow down resulting in a larger Doppler-shift and {\it vice versa}. Consequently, the $\gamma$-ray line shape is sensitive to the lifetime of the excited nucleus.
In addition to the lifetime, several experimental effects also influence the $\gamma$-ray line shape. These are discussed in Section~\ref{sec:lineshape}.%

For very short-lived excited states the average speed reduction is roughly
$\delta \beta \approx 0.32\times 10^{-3} \times (\tau/A) \times (\textrm{d}E/\textrm{d}x)$,
where $\tau$ is the mean lifetime in units of fs, $A$ is the mass number of the excited recoil, and $\textrm{d}E/\textrm{d}x$ is the stopping power in units of MeV/$\mu$m. 
Three important observations can be made from this formula: 
First, a dense medium with a high stopping power gives the highest sensitivity. This makes Au an ideal stopping medium. %
Second, the stopping power must be known to extract the lifetime. At the speeds relevant to the present study ($\beta = 0.059$--0.082) the energy loss of Mg ions in Au is dominated by electronic stopping, and the stopping powers have been determined experimentally by Forster {\it et al.}~\cite{forster76} to a precision of $\pm4\%$. Approximately, $\textrm{d}E/\textrm{d}x \approx 8.5$~MeV/$\mu$m.  
Third, lifetimes shorter than 1~fs will be very difficult to measure even in an idealized experiment. Assuming $\tau=1$~fs one obtains $\delta \beta \approx 1.2\times 10^{-4}$, which implies a shift of only 0.9~keV for the $7787\,\textrm{keV}\rightarrow 451\,\textrm{keV}$ transition. 
%

\section{Setup and Procedure}\label{sec:setup}

\begin{figure*}
\includegraphics[width=0.8\linewidth, angle=0, clip=true, trim= 0 0 0 0]{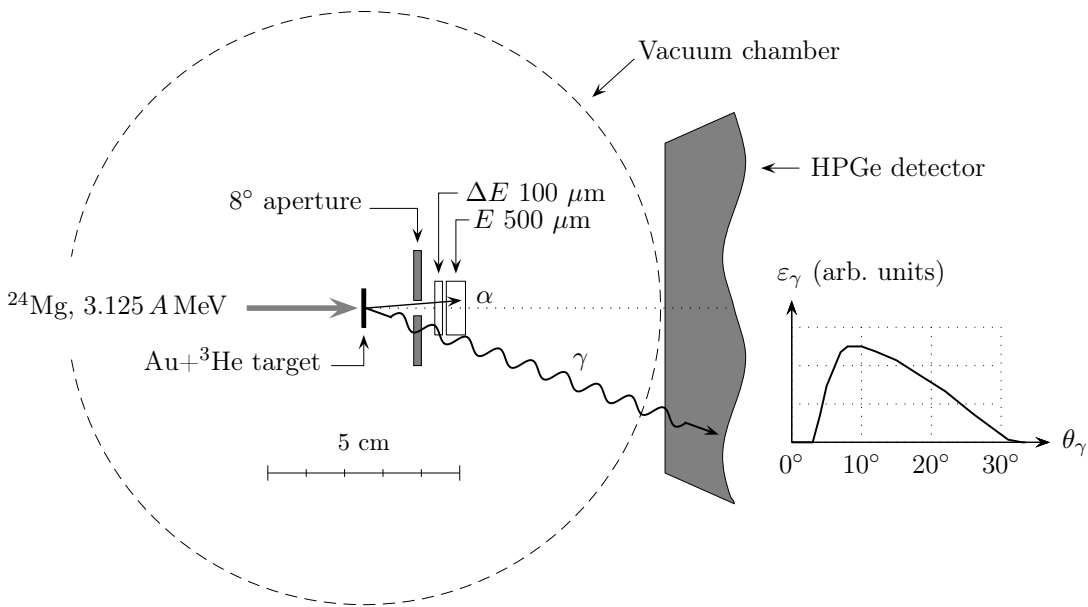}%
\caption{Schematic drawing of the experimental setup. The setup also includes a cooling system and a collimator not shown on the drawing. The collimator consists of two circular apertures with diameters of 2.5 and 3.0~mm placed at distances of 73 and 49~mm, respectively, upstream from the target. $\varepsilon_{\gamma}$ is the energy- and angle-dependent detection efficiency of the HPGe detector, here shown for a fixed energy of $E_{\gamma}=8$~MeV.}
\label{fig:setup}
\end{figure*}

\begin{figure}
\includegraphics[width=1.0\linewidth, angle=0, clip=true, trim= 0 0 0 0]{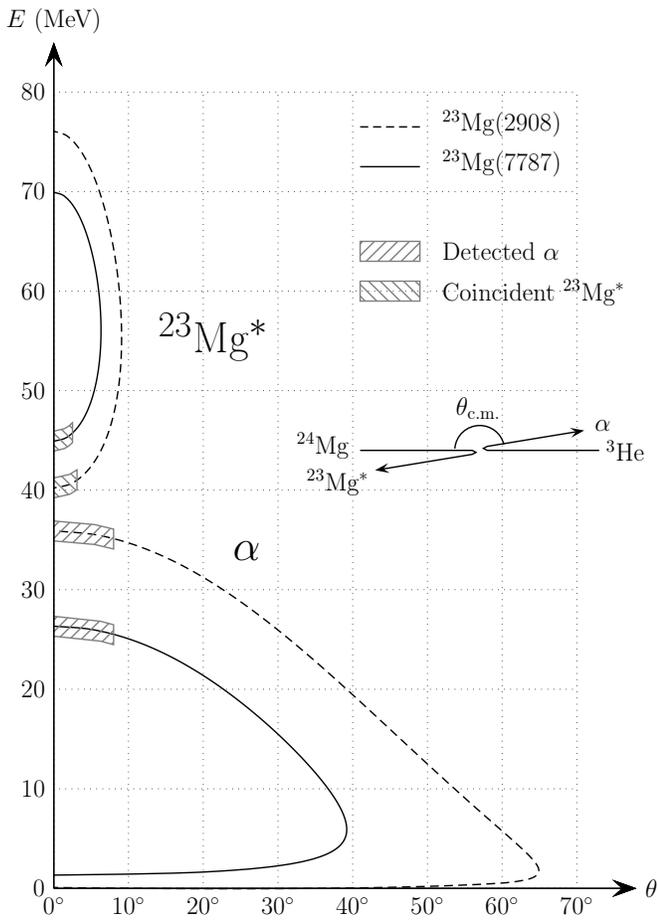}%
\caption{\label{fig:kincur}Kinematic curves for the $^3$He$(^{24}$Mg$,\alpha)$ reaction at a beam energy of 3.125$\,A\,$MeV. Hatched regions indicate detected $\alpha$ particles and associated $^{23}$Mg$^{\ast}$ recoils.}
\end{figure}

The experiment was carried out at the ISAC-II facility of TRIUMF. A schematic drawing of the setup is shown in Fig.~\ref{fig:setup}. A similar setup has been successfully employed in previous experiments at TRIUMF to measure lifetimes of excited states in $^{19}$Ne~\cite{kanungo06, mythili08} and $^{15}$O~\cite{galinski2014}. %
\begin{figure*}
\includegraphics[width=0.8\linewidth, angle=0, clip=true, trim= 0 0 0 0]{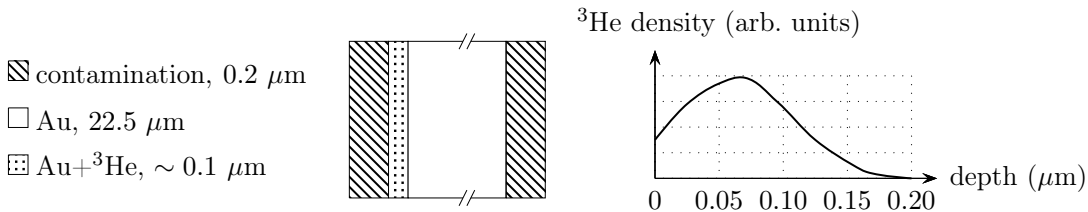}%
\caption{Target composition and $^3$He depth profile predicted by SRIM~\cite{srim}. The surface contamination resulted from condensation of residual gas and water vapor.}
\label{fig:target}
\end{figure*}
The $^{24}$Mg beam is accelerated to 3.125$\,A\,$MeV corresponding to $\beta = 0.082$ and directed onto a $^3$He-implanted Au foil. Excited states in $^{23}$Mg are populated via $^3$He$(^{24}$Mg$,\alpha)$ with a $Q$-value of 4.05~MeV. Kinematic curves are shown in Fig.~\ref{fig:kincur}. Because the experiment is performed in inverse kinematics, the recoils are strongly focused in the forward direction with speeds in the range $\beta = 0.059$--0.065. %
The $\alpha$ particles are detected using a $\Delta E$-$E$ telescope placed downstream of the target, consisting of two Si surface-barrier detectors with an active area of 150~mm$^2$ and thicknesses of 100~$\mu$m and 500~$\mu$m. The target is sufficiently thick to stop the beam and the recoils, thus preventing damage of the detectors, while thin enough to allow transmission of the $\alpha$ particles. A 4~mm diameter circular aperture placed at a distance of 14.2~mm downstream from the target limits the angular acceptance of the $\Delta E$-$E$ telescope to 0--8$^{\circ}$. A $\gamma$-ray detector consisting of four separate HPGe crystals is situated 7.83~cm downstream of the target and covers angles up to $\approx 32^{\circ}$. The placement of the detectors at $0^{\circ}$ serves to maximize the Doppler shift. %
By gating on the $\alpha$ particles we suppress competing channels such as $n$, $p$, $2p$, $d$, $n+p$, $^{3}$He, $d+p$, and $n+2p$, resulting in a significantly cleaner $\gamma$-ray spectrum. By gating on the $\alpha$ particles we also reduce the spread in recoil velocities (magnitude and angle) and thus the kinematic broadening of the $\gamma$-ray lines. %
The target was prepared at the Universit\'e de Montreal by directing a 30~keV $^3$He beam onto $22.6\pm 0.2$~$\mu$m thick Au foil. In this way the $^3$He ions were implanted at shallow depth with a number density of $6.6\times 10^{17}$~cm$^{-2}$. The depth profile predicted by SRIM~\cite{srim} is shown in Fig.~\ref{fig:target}. It peaks at 0.07~$\mu$m, but is quite broad with a tail that extends 0.15~$\mu$m into the Au foil. The thickness of the Au foil was determined using a precision mass balance. %

The integrated beam time was 5 days. Typical beam intensities were 1--$2\times 10^{10}$~ions/s. The target temperature was kept at $-80^{\circ}$ using a LN$_2$ cooling system, see Refs.~\cite{kanungo06, mythili08, galinski2014} for details. This is necessary to prevent loss of $^3$He due to beam-induced heating of the target. %
The elastic scattering rate was used for on-line monitoring of the $^3$He content of the target. No indication of $^3$He loss was seen.

\begin{figure}
\includegraphics[width=0.96\linewidth, angle=0, clip=true, trim= 20 40 20 40]{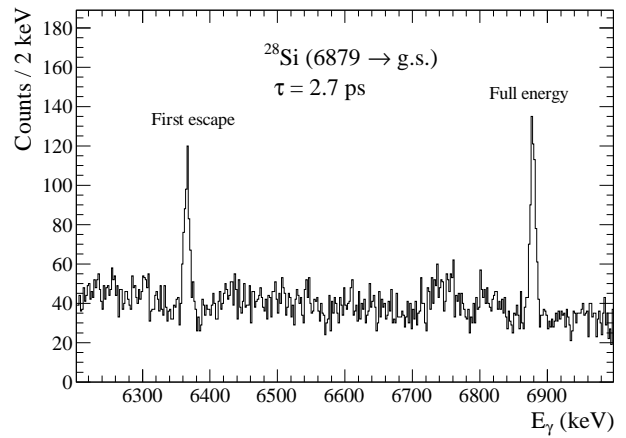}%
\caption{The unshifted $\gamma$-ray line due to the $6879\,\textrm{keV}\rightarrow\,\textrm{g.s.}$\ transition in $^{28}$Si.}
\label{fig:si28}
\end{figure}

During the initial pump-down residual gas and water vapor was inadvertently condensed on the surface of the cooled target due to poor vacuum. The thickness of the condensation layer has been estimated to be 0.2~$\mu$m of water equivalent based on the observed energy of the main $^3$He and $\alpha$-particle groups. As a result the beam energy was degraded by 0.25~MeV. Reactions between the beam and the surface contaminants contribute to the background in the $\gamma$-ray spectrum, but serendipitously also provide us with two unshifted $\gamma$-ray lines at high energy, namely, the 6877.88(8)~keV line from $^{28}$Si and its single-escape peak, which enable us to determine the energy calibration and the intrinsic detector resolution very precisely in the energy region of interest. 
(The $\gamma$-ray energy given here differs slightly from the $\gamma$-ray energy of 6877.0~keV given in the NDS evaluation~\cite{nuclear_data_sheets_A28}. The $\gamma$-ray energy given here was calculated from the excitation energies given in the NDS evaluation, corrected for the nuclear recoil energy.)
The 6877.88~keV line and its single-escape peak are shown in Fig.~\ref{fig:si28}. The uncertainty on the energy calibration is $\pm 1$~keV at 7.8~MeV. Gain variations during the experiment were below $\pm 0.5$~keV. 
The yield of the 6877.88~keV line relative to the elastic scattering rate was constant during the experiment, confirming that the condensation took place only during the initial pump-down of the chamber.

\section{Line Shape}\label{sec:lineshape}

Several experimental effects contribute to the $\gamma$-ray line shape. In order to extract reliable lifetimes these effects must be accurately and precisely modeled. Below, we list the effects and  estimate their contribution to the relative spread in $\gamma$-ray energy, $\delta E_{\gamma} / E_{\gamma}$, at $E_{\gamma}=7.8$~MeV. The estimates are rather crude and merely serve to provide an idea of the relative importance of the various effects. The quantitative treatment is discussed in Section~\ref{sec:simulation}.

\begin{description}%
\item[Intrinsic resolution] The intrinsic resolution of the $\gamma$-ray detector is reflected in the line shape of unshifted peaks. At the highest $\gamma$-ray energies the line shape is found to be well described by a Gaussian; at the lowest $\gamma$-ray energies two additional components, a skewed Gaussian and a smoothed step function, must be included to obtain an adequate fit, see Ref.~\cite{galinski2014} for details. Fits to the unshifted $\gamma$-ray lines due to the $279\,\textrm{keV}\rightarrow\,\textrm{g.s.}$\ transition in $^{197}$Au ($\tau=26.8$~ps) and the $6879\,\textrm{keV}\rightarrow\,\textrm{g.s.}$\ transition in $^{28}$Si ($\tau=2.7$~ps) are shown in Fig.~\ref{fig:response}. The FWHM of the Gaussian component is found to obey the relation, $\textrm{FWHM} = 3.3\,\textrm{keV}\times (E_{\gamma}/$MeV$)^{0.52}$. %
This gives $\delta E_{\gamma} / E_{\gamma} \approx 1.2\times 10^{-3}$ at $E_{\gamma}=7.8$~MeV.
\begin{figure*}
\includegraphics[width=0.9\linewidth, angle=0, clip=true, trim= 0 15 25 25]{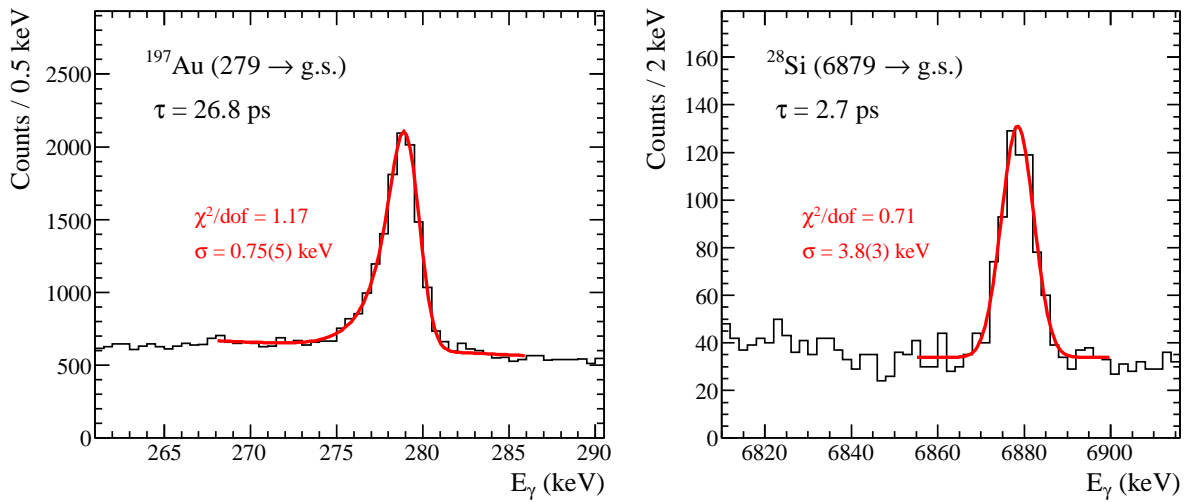}%
\caption{(Color online) Fits to the unshifted $\gamma$-ray lines due to the $279\,\textrm{keV}\rightarrow\,\textrm{g.s.}$\ transition in $^{197}$Au and the $6879\,\textrm{keV}\rightarrow\,\textrm{g.s.}$\ transition in $^{28}$Si.}
\label{fig:response}
\end{figure*}
\item[Angular acceptance] The finite angular acceptance of the detectors results in kinematic broadening of the Doppler-shifted $\gamma$-ray lines. Accurate modeling of this effect requires knowledge of the detector geometry and the angular dependence of the $\gamma$-detection efficiency. In principle knowledge of the angular distribution of the $\alpha$ particles and the $\alpha$-$\gamma$ angular-correlation function is also required, but their effect on the line shape is rather small due to the limited angular acceptance. %
The finite acceptance of the $\Delta E$-$E$ telescope causes a spread of $\delta E_{\gamma} / E_{\gamma} \approx 0.6\times 10^{-3}$, while the finite acceptance of the $\gamma$-ray detector causes a spread of $\delta E_{\gamma} / E_{\gamma} \approx 0.6\times 10^{-2}$.
\item[Beam emittance] The finite emittance of the beam also contributes to the line shape. The energy spread translates directly into a spread in speed and thus a spread in the Doppler shift. In the present experiment the energy spread of the beam was $0.2$\% (FWHM), resulting in a spread of $\delta E_{\gamma} / E_{\gamma} \approx 0.6\times 10^{-4}$. %
The divergence and transverse size of the beam influences the line shape in a subtler way by allowing the detection of $\alpha$ particles with scattering angles larger than the $8^{\circ}$ acceptance of the circular aperture in front of the $\Delta E$-$E$ telescope.
The divergence of the beam can safely be neglected.
The transverse size is constrained by the 2.5~mm diameter circular aperture placed at a distance of 73~mm upstream from the target. Visual inspection of the target at the end of the experiment revealed a burn mark with a diameter of 2--3~mm, consistent with the size of the aperture. A beam diameter of 2.5~mm thus appears realistic, and we conclude that $\alpha$ particles with scattering angles as large as 13$^{\circ}$ reached the detector. We estimate the spread caused by this effect to be $\delta E_{\gamma} / E_{\gamma} \approx 0.9\times 10^{-3}$.

%
\item[\texorpdfstring{$^3$He}{3He} depth profile] Since a reaction can take place at any depth inside the $^3$He-implanted layer, the beam-energy resolution is fundamentally limited by the energy loss inside this layer. The maximum energy loss, which occurs if the reaction takes place at the bottom of the $^3$He-implanted layer near a depth of 0.15~$\mu$m, is 1.0~MeV. This results in a spread of $\delta E_{\gamma} / E_{\gamma} \approx 0.4\times 10^{-3}$. 
%
%
%
\item[Multiple scattering] Angular deflection due to multiple scattering in the target influences the line shape in the same way as the finite beam emittance by allowing the detection of $\alpha$ particles with scattering angles larger than the $8^{\circ}$ acceptance. 
The half-width of the angular distribution ($\alpha_{1/2}$ in Eq.~(\ref{eq:angular})) may be estimated from Eq.~(20) in Ref.~\cite{anne88}, though the stated range of applicability (20--90$\,A\,$MeV) is somewhat above the range of speeds probed in the present study (5--10$\,A\,$MeV for the $\alpha$ particles and 1--2$\,A\,$MeV for the $^{23}$Mg recoils). For the $\alpha$ particles the half-width after passage through the Au foil is found to vary between 1.8$^{\circ}$ and 3.6$^\circ$, the precise value depending on the energy, resulting in a typical spread of $\delta E_{\gamma} / E_{\gamma} \approx 1.0\times 10^{-3}$. The $^{23}$Mg recoils are also deflected, but less so: the half-width after a distance of 0.36~$\mu$m in the Au foil (the distance covered in approximately 20~fs) is found to vary between 0.6$^\circ$ and 1.2$^\circ$, the precise value again depending on the energy. The lateral deflection caused by multiple scattering can safely be neglected.

\item[Energy-loss straggling] The energy-loss straggling may be estimated using the semi-empirical formula of Ref.~\cite{guillemaud-mueller86}. Using the same numbers as above (40~MeV, 0.36~$\mu$m) the energy loss of the $^{23}$Mg recoils is determined to be 3.4~MeV and the straggling to be $0.23$~MeV (FWHM). This results in a spread of $\delta E_{\gamma} / E_{\gamma} \approx 1.8\times 10^{-4}$.

\end{description}

The experimental effects contributing to the $\gamma$-ray line shape are summarized in Table~\ref{tb:lineshape}.
\begin{table}
\caption{\label{tb:lineshape} Experimental effects contributing to the $\gamma$-ray line shape and their relative importance for the $7786\,\textrm{keV}\rightarrow 451\,\textrm{keV}$ transition in $^{23}$Mg compared to the shift caused by a lifetime of 1~fs.}
\begin{ruledtabular}
\renewcommand{\thefootnote}{\alph{footnote}}
\begin{tabular}{lcc}
Experimental effect  &  $\delta E_{\gamma} / E_{\gamma}$~(\textperthousand) \\
\hline
Intrinsic resolution         &  1.3\phantom{0$^a$} \\
$\Delta E$-$E$ acceptance    &  0.6\phantom{0$^a$} \\ 
HPGe acceptance              &  6\phantom{.00$^a$} \\
Beam-energy spread           &  0.06\phantom{$^a$} \\
Beam transverse size         &  0.9\phantom{0$^a$} \\
$^3$He depth profile         &  0.4\phantom{0$^a$} \\ 
Multiple scattering          &  1.0\footnotemark[1]\phantom{0}  \\  
Energy-loss straggling       &  0.18\footnotemark[1]  \\
\hline
Shift caused by $\tau=1$~fs  &  0.12\phantom{$^a$} \\
\hline
\end{tabular}
\footnotetext[1]{Assuming $\tau=20$~fs. For shorter lifetimes the effect is smaller.}
\end{ruledtabular}
\end{table}

\section{Monte Carlo Simulation}\label{sec:simulation}

\begin{figure*}
\includegraphics[width=0.7\linewidth, angle=0, clip=true, trim= 0 0 0 0]{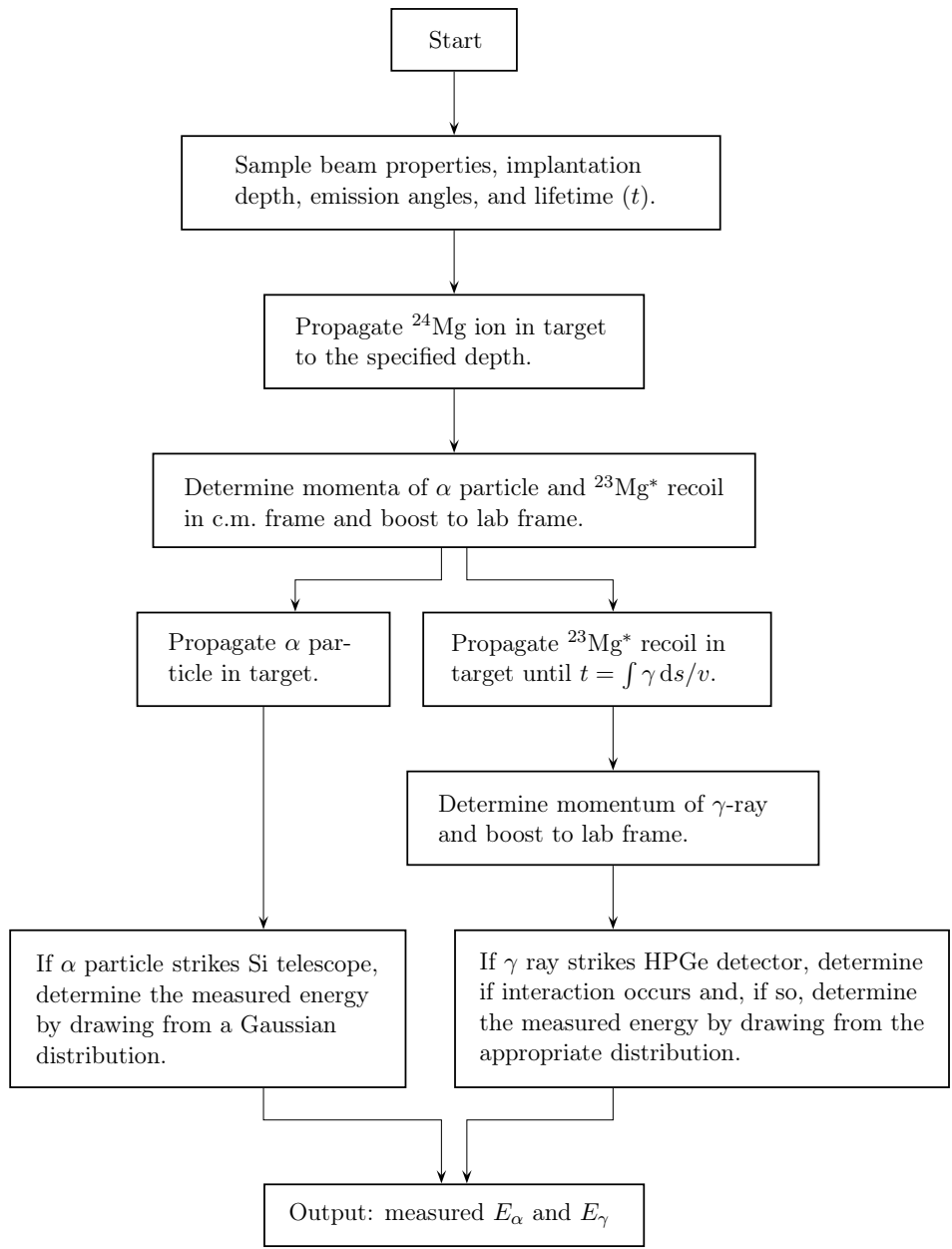}
\caption{Flowchart of the Monte Carlo simulation program. See text for details.}
\label{fig:flowchart}
\end{figure*}

A quantitative treatment of the experimental effects discussed in the preceding section is most easily accomplished with a Monte Carlo simulation. A simulation program has been written in the language C++. The program makes use of functionalities provided by the data analysis framework ROOT~\cite{brun1997}. A flowchart of the program is shown in Fig.~\ref{fig:flowchart}. The program may be obtained at \url{https://github.com/oliskir/SimDSAM.git}.

We begin by sampling the properties of the beam particle, {\it i.e.}, its energy and transverse displacement from the beam axis, the depth of the implanted $^3$He atom in the target, the emission angles of the $\alpha$ particle in the c.m.\ frame, the emission angles of the $\gamma$ ray in the rest frame of the $^{23}$Mg$^{\ast}$ recoil, and the lifetime of the excited state ($t$). For the beam we assume a Gaussian energy distribution with a spread of 0.2\% (FWHM) and a uniform circular transverse profile with a diameter of 2.5~mm. For the $^3$He implantation depth we assume the profile determined by SRIM~\cite{srim} shown in Fig.~\ref{fig:target}. For the emission angles we generally assume isotropic distributions. The error introduced by this simplifying assumption has been estimated by trying a few different realistic anisotropic distributions and has been found to be negligible compared to other errors for all the transitions considered, except the $2908\,\textrm{keV}\rightarrow\,\textrm{g.s.}$\ transition where it gives an error of $\pm 2$~fs. Finally, the lifetime is drawn from the standard exponential-decay distribution. %

Next, we propagate the beam particle in the target until it reaches the specified depth. This involves three tasks: First, the energy loss is calculated by numerical integration of the stopping powers (cf.~Section~\ref{ref:stopping}) assuming that the particle travels along a straight path. (This assumption implies a slight underestimate of the energy loss which, however, is negligible at the energies relevant to the present study.) Second, the angular deflection due to multiple scattering, $\alpha$, is drawn from the distribution (Eq.~(8) in Ref.~\cite{anne88})
\begin{equation}\label{eq:angular}
f( \alpha )  \; \propto \;  \sin \left( \alpha \right)  \,  \exp \left( - \ln 2 \, \frac{\alpha^2}{\alpha_{1/2}^2} \right) \; ,
\end{equation}
where $\alpha_{1/2}$ is the half-width given by Eq.~(20) in Ref.~\cite{anne88}. Note that for small deflections $\sin \left( \alpha \right) \approx \alpha$, implying a ``universal'' distribution of the form $f(u) \propto u \exp(-u^2)$ with $u = \sqrt{\ln 2} \, \alpha / \alpha_{1/2} $. %
Third, the energy-loss straggling is sampled from a Gaussian distribution with the standard deviation given by the semi-empirical formula of Ref.~\cite{guillemaud-mueller86}. %

We determine the momenta of the $\alpha$ particle and the excited $^{23}$Mg$^{\ast}$ recoil in the centre of mass (c.m.)\ frame, using the already sampled emission angles, and boost to the laboratory (lab) frame, always using relativistic kinematics. %
We then propagate the $\alpha$ particle and the $^{23}$Mg$^{\ast}$ recoil in the target, the former until it exits, the latter until it decays, {\it i.e.}, until $t = \int \gamma \, \textrm{d}s/v$, where $v=c\beta$ is the instantaneous speed of the recoil, $\gamma=(1-\beta^2)^{-1/2}$, and $\textrm{d}s$ is an infinitesimal step along the trajectory of the recoil. %
We then determine the momentum of the emitted $\gamma$ ray, again using the already sampled emission angles, and boost to the lab frame. %

Finally, we check if the $\alpha$ particle and the $\gamma$ ray strike the detectors and sample the intrinsic resolution of each detector. For the $\gamma$-ray detector we also use the energy- and angle-dependent detection efficiency, $\varepsilon_{\gamma}(E_{\gamma},\theta_{\gamma})$, determined by a GEANT4 simulation~\cite{svensson2014} and shown in Fig.~\ref{fig:setup} for $E_{\gamma}=8$~MeV. (In practice, simulations were performed for $E_{\gamma}=0.1,\,0.5,\,1,\,2,\,3\, \dots ,10$~MeV and the efficiency at intermediate energies was determined by cubic-spline interpolation.) %

\section{Stopping Powers}\label{ref:stopping}
Inside the $^3$He-implanted layer the stopping power is different from that in pure Au. The presence of the $^3$He atoms acts to increase the stopping power, but also causes the region to swell which lowers the density and thus the stopping power. This must be taken into account when the energy loss is calculated. %
Letting $c(x)$ denote the local relative concentration of $^3$He at depth $x$ in the target, the local swelling is given by $\Delta V/V = A \, c(x)$ where $A = 0.75 \pm 0.25$~\cite{alexander81}. 
We take $c(x)$ to be the $^3$He depth profile predicted by SRIM~\cite{srim} normalized to the total number of implanted $^3$He atoms, $6.6\times 10^{17}$~cm$^{-2}$. As seen in Fig.~\ref{fig:target} the concentration peaks at a depth of $x_0=0.07$~$\mu$m, reaching a maximum value of $c(x_0)=1.05$.
The stopping power may then be calculated as
\begin{equation}
\frac{\textrm{d}E}{\textrm{d}x} 
= \frac{\rho_{\textrm{Au}}}{1+A\;c(x)} 
\left[ \;
\left( \frac{\textrm{d}E}{\rho\, \textrm{d}x} \right)_{\textrm{Au}}
+ \frac{3}{197} \, c(x) \, 
\left( \frac{\textrm{d}E}{\rho\, \textrm{d}x} \right)_{^{3}\textrm{He}} 
\; \right]
\end{equation}
where $\rho_{\textrm{Au}}=19.30$~g/cm$^3$ is the density of gold and $(\textrm{d}E / \rho\, \textrm{d}x)_{\textrm{Au}}$ and $(\textrm{d}E / \rho\, \textrm{d}x)_{^3\textrm{He}}$ are the stopping powers in Au and $^3$He in units of MeV/(g/cm$^2$). For Au we use the experimental stopping powers of Ref.~\cite{forster76}, measured to a precision of $\pm4\%$. For $^3$He we use SRIM stopping powers~\cite{srim}. In both cases we interpolate between the tabulated values using cubic-spline interpolation.

\section{Data Analysis}

\begin{figure}
\includegraphics[width=0.95\linewidth, angle=0, clip=true, trim= 0 45 20 40]{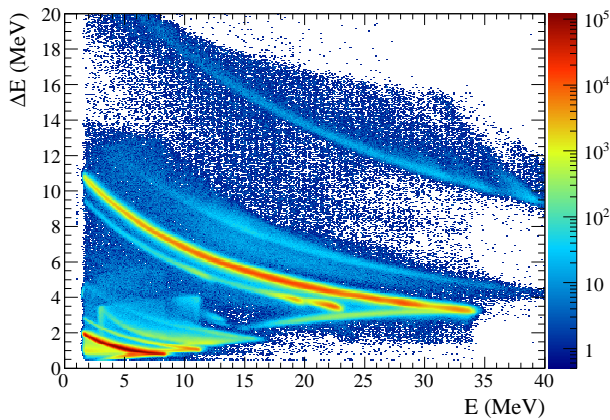}%
\caption{(Color online) $\Delta E$-$E$ plot used to identify the light reaction products. Each banana-shaped band represents a reaction product or a coincidence of two reaction products. From top to bottom: $\alpha+\alpha$, $p+\alpha$, $\alpha$, $^3$He, $p+d$, $p+p$, $d$, and $p$.}
\label{fig:dee}
\end{figure}

As shown in Fig.~\ref{fig:dee}, the differential energy-loss data from the $\Delta E$-$E$ telescope allows us to cleanly separate $\alpha$ particles from other light ejectiles, such as $p$, $d$, and $^3$He. The reaction channel of interest, $(^3$He$,\alpha)$, produces $\alpha$ particles with a well-defined energy at $0^{\circ}$ degrees, providing unambiguous determination of the excitation energy of the $^{23}$Mg$^{\ast}$ recoil. Experimental effects smear out the measured energy, resulting in an $\alpha$-particle energy resolution of 1~MeV (FWHM), which translates into an excitation-energy resolution of about 0.5~MeV (FWHM).
A standard $\alpha$ source, consisting of $^{239}$Pu, $^{241}$Am, and $^{244}$Cm, was used to calibrate the Si detectors. 

\begin{figure}
\includegraphics[width=0.95\linewidth, angle=0, clip=true, trim= 20 30 70 70]{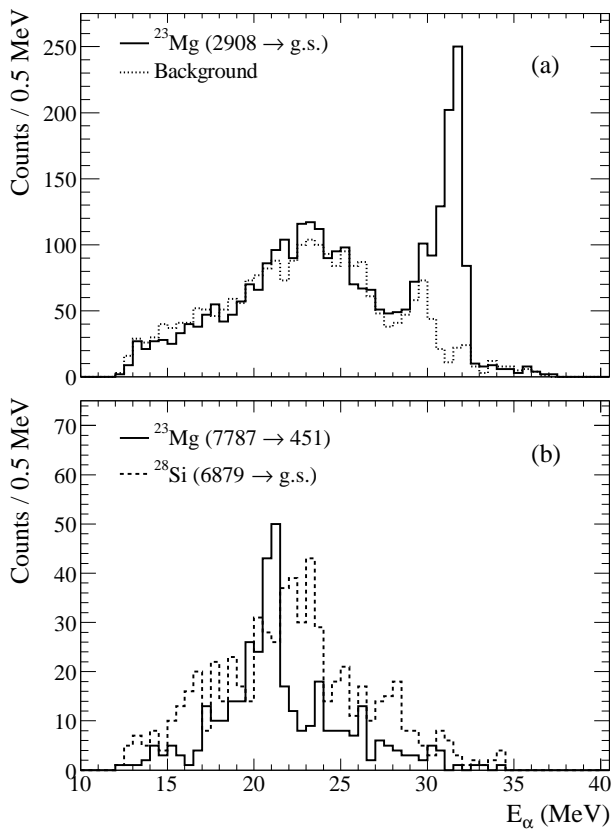}%
\caption{$\alpha$-Particle spectra obtained by gating on the $2908\,\textrm{keV}\rightarrow\,\textrm{g.s.}$\ and $7787\,\textrm{keV}\rightarrow 451\,\textrm{keV}$ transitions in $^{23}$Mg and the $6878\,\textrm{keV}\rightarrow\,\textrm{g.s.}$\ transition in $^{28}$Si. The background spectrum shown by the dashed line in (a) has been obtained by gating on an interval of $\gamma$-ray energies just above the $\gamma$-ray peak.}
\label{fig:alphaspec}
\end{figure}

By gating on the Doppler-shifted $\gamma$-ray line of the $2908\,\textrm{keV}\rightarrow\,\textrm{g.s.}$\ transition in $^{23}$Mg, one obtains the $\alpha$-particle spectrum drawn by the solid line in Fig.~\ref{fig:alphaspec}~(a), where a clear peak is seen at $E_{\alpha}\approx 32$~MeV. When the energy loss of the $\alpha$ particles in the Au foil is taken into account ($\approx 4$~MeV), the peak energy is found to be consistent with the excitation energy of the 2908~keV state, cf.~Fig.~\ref{fig:kincur}. %
The $\alpha$-particle spectrum obtained by gating on the Doppler-shifted $\gamma$-ray line of the $7787\,\textrm{keV}\rightarrow 451\,\textrm{keV}$ transition is shown by the solid line in Fig.~\ref{fig:alphaspec}~(b). Again a clear peak is seen, this time at $E_{\alpha}\approx 21$~MeV, an energy that is consistent with the excitation energy of the 7787~keV state. In contrast, no narrow peak is seen in the $\alpha$-particle spectrum obtained by gating on the unshifted $\gamma$-ray line of the $6878\,\textrm{keV}\rightarrow\,\textrm{g.s.}$\ transition in $^{28}$Si. This strongly suggests that the $^{28}$Si$^{\ast}$ recoil is produced in a fusion-evaporation reaction, such as $^{24}$Mg$+{}^{12}$C$\,\rightarrow 2\alpha+{}^{28}$Si$^{\ast}$ or $^{24}$Mg$+{}^{16}$O$\,\rightarrow 3\alpha+{}^{28}$Si$^{\ast}$. The $\gamma$-gated $\alpha$-particle spectrum may thus provide important clues to the origin of unidentified $\gamma$-ray lines. %

By inverting the procedure, we may generate $\alpha$-gated $\gamma$-ray spectra with rather tight cuts on the excitation energy in $^{23}$Mg. Thus, feeding from higher-lying states is efficiently suppressed, which simplifies the line-shape analysis considerably. %
The $\alpha$-gated $\gamma$-ray spectra of the 2908~keV and 7787~keV states are shown in Fig.~\ref{fig:lifetime03} and Fig.~\ref{fig:lifetime15},\ref{fig:lifetime11},\ref{fig:lifetime12},\ref{fig:lifetime14},\ref{fig:lifetime15}. The gate on the $\alpha$-particle energy was 2.4~MeV wide in both cases. Simulated line shapes have been fitted to the experimental data for three different mean lifetimes, using the method of maximum likelihood~\cite{baker1984}. The only free parameters are the normalization, two or three parameters to describe the background, and an energy shift, which is allowed to vary within $\pm (\sigma_{\textrm{tab}} + \sigma_{\textrm{cal}})$ where $\sigma_{\textrm{tab}}$ is the uncertainty on the tabulated energy (typically $1\sigma$) and $\sigma_{\textrm{cal}}$ is the $1\sigma$ uncertainty on the energy calibration. %
The background was modeled as a constant plus a smoothed step function in the case of the $2908\,\textrm{keV}\rightarrow\,\textrm{g.s.}$\ transition, and as linear function plus a smoothed step function in the case of the $7787\,\textrm{keV}\rightarrow 451\,\textrm{keV}$ transition. The $\chi^2/$dof is given for each fit (the number enclosed in parentheses).
%
%
The mean lifetime of the 2908~keV state is determined to be $15\pm 3$~fs, while an upper limit of 12~fs is obtained for the 7787~keV state at the 95\% C.L. The overall error of $\pm 3$~fs on the mean lifetime of the 2908~keV state includes a systematic uncertainty of $\pm 2$~fs from the angular distribution (cf.\ Section~\ref{sec:simulation}), which has been added in quadrature to the statistical uncertainty from the fit.
We note that the fit is not too sensitive to the constraints imposed on the energy shift. If the constraints are completely removed, the preferred value for the mean lifetime of the 2908~keV state is reduced slightly to 12~fs, and the upper limit on the mean lifetime of the 7787~keV state is increased slightly to 14~fs.

\section{Results and Discussion}

\begin{figure}
\includegraphics[width=0.95\linewidth, angle=0, clip=true, trim= 20 30 60 50]{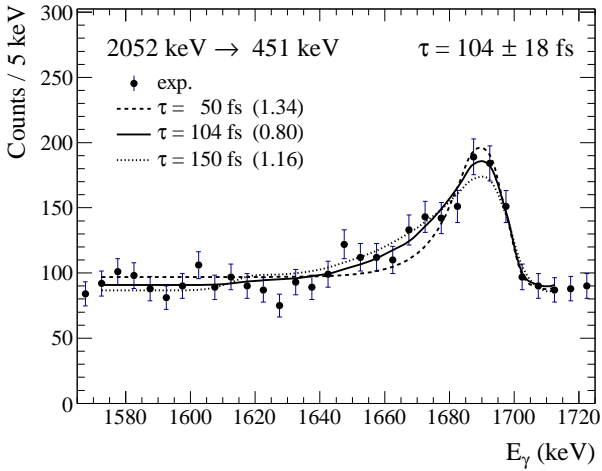}%
\caption{\label{fig:lifetime01}Line-shape analysis of the $\alpha$-gated, doppler-shifted, $\gamma$-ray line of the $2052\,\textrm{keV}\rightarrow\,451\,\textrm{keV}$ transition in $^{23}$Mg. The curves show simulated line shapes, which have been fitted to the experimental data (see text for details). The numbers in parentheses give the $\chi^2/$dof of each fit.}
\end{figure}

\begin{figure}
\includegraphics[width=0.95\linewidth, angle=0, clip=true, trim= 20 30 60 50]{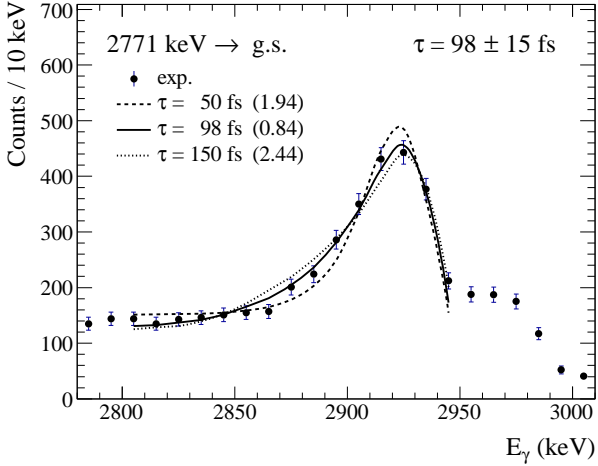}%
\caption{\label{fig:lifetime02}The $2771\,\textrm{keV}\rightarrow\,\textrm{g.s.}$\ transition.}
\end{figure}

\begin{figure}
\includegraphics[width=0.95\linewidth, angle=0, clip=true, trim= 20 30 60 50]{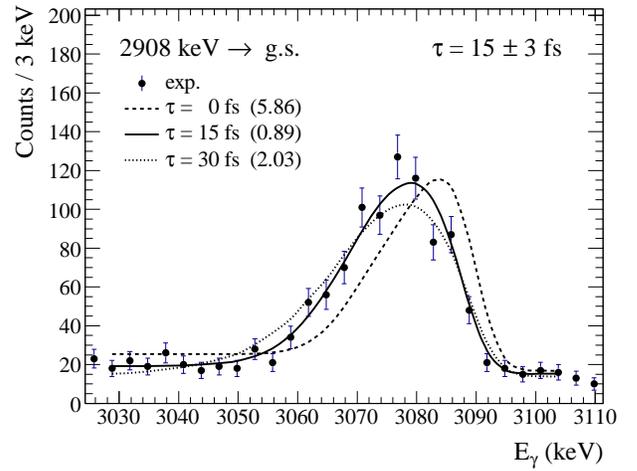}%
\caption{\label{fig:lifetime03}The $2908\,\textrm{keV}\rightarrow\,\textrm{g.s.}$\ transition.}
\end{figure}

\begin{figure}
\includegraphics[width=0.95\linewidth, angle=0, clip=true, trim= 20 30 60 50]{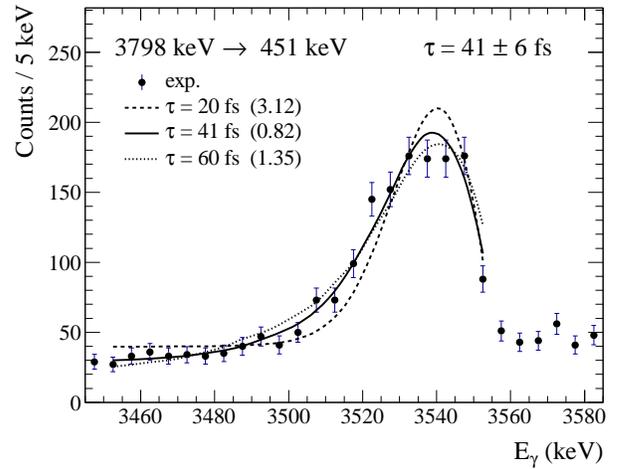}%
\caption{\label{fig:lifetime05}The $3798\,\textrm{keV}\rightarrow\,451\,\textrm{keV}$\ transition.}
\end{figure}

\begin{figure}
\includegraphics[width=0.95\linewidth, angle=0, clip=true, trim= 20 30 60 50]{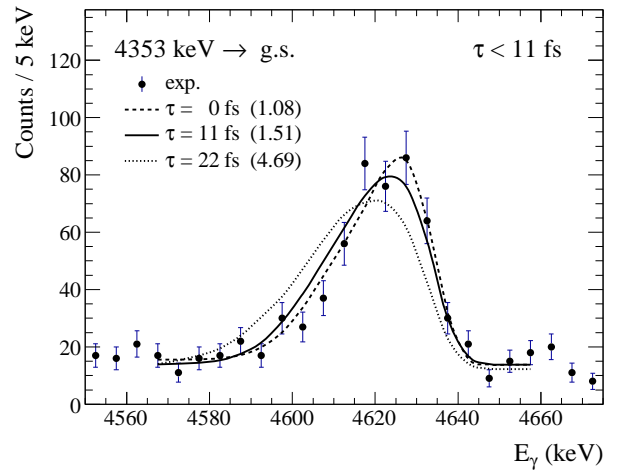}%
\caption{\label{fig:lifetime06}The $4353\,\textrm{keV}\rightarrow\,\textrm{g.s.}$\ transition.}
\end{figure}

\begin{figure}
\includegraphics[width=0.95\linewidth, angle=0, clip=true, trim= 20 30 60 50]{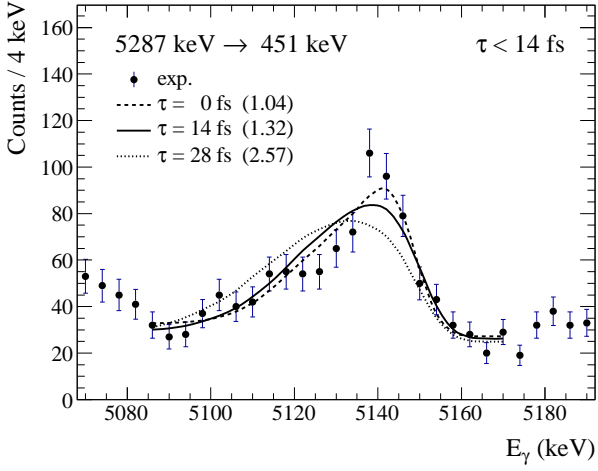}%
\caption{\label{fig:lifetime07}The $5287\,\textrm{keV}\rightarrow\,451\,\textrm{keV}$\ transition.}
\end{figure}

\begin{figure}
\includegraphics[width=0.95\linewidth, angle=0, clip=true, trim= 20 30 60 50]{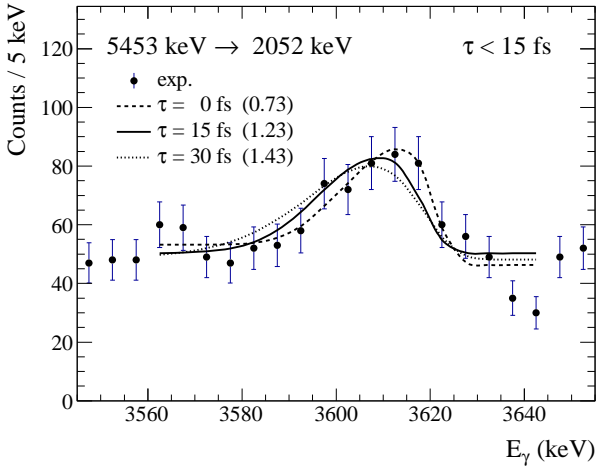}%
\caption{\label{fig:lifetime09}The $5453\,\textrm{keV}\rightarrow\,2052\,\textrm{keV}$\ transition.}
\end{figure}

\begin{figure}
\includegraphics[width=0.95\linewidth, angle=0, clip=true, trim= 20 30 60 50]{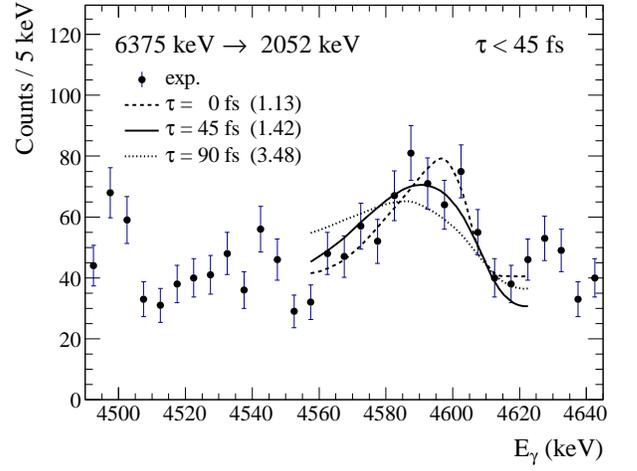}%
\caption{\label{fig:lifetime11}The $6375\,\textrm{keV}\rightarrow\,2052\,\textrm{keV}$\ transition.}
\end{figure}

\begin{figure}
\includegraphics[width=0.95\linewidth, angle=0, clip=true, trim= 20 30 60 50]{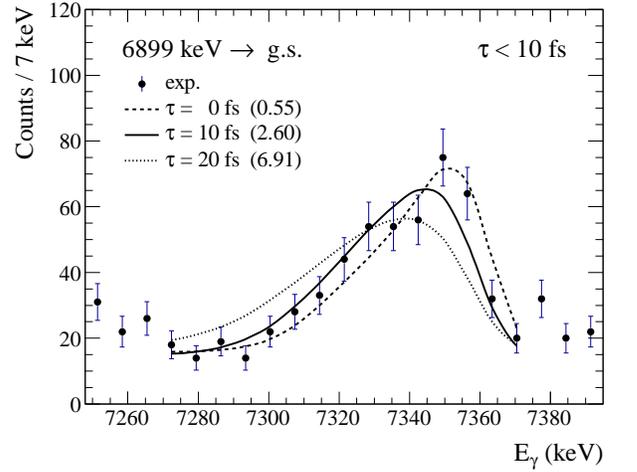}%
\caption{\label{fig:lifetime12}The $6899\,\textrm{keV}\rightarrow\,\textrm{g.s.}$\ transition.}
\end{figure}

\begin{figure}
\includegraphics[width=0.95\linewidth, angle=0, clip=true, trim= 20 30 60 50]{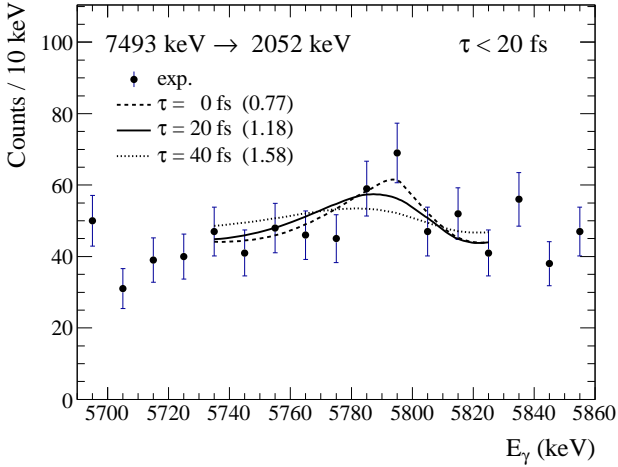}%
\caption{\label{fig:lifetime14}The $7493\,\textrm{keV}\rightarrow\,2052\,\textrm{keV}$\ transition.}
\end{figure}

\begin{figure}
\includegraphics[width=0.95\linewidth, angle=0, clip=true, trim= 20 30 60 50]{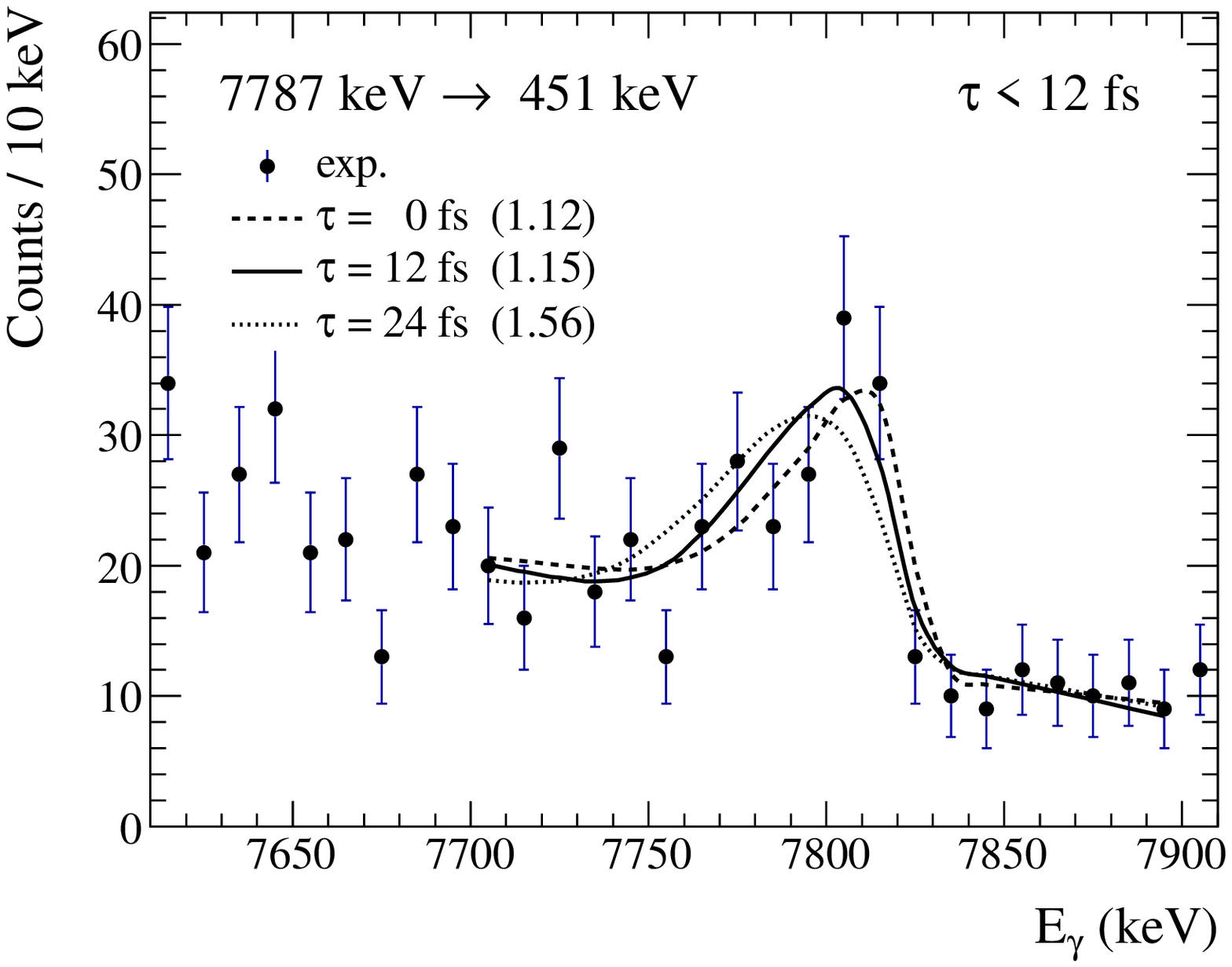}%
\caption{\label{fig:lifetime15}The $7787\,\textrm{keV}\rightarrow\,451\,\textrm{keV}$\ transition.}
\end{figure}

The $\gamma$-ray energies and mean lifetimes determined from the present study are listed in Table~\ref{tb:results}, and the experimental $\gamma$-ray spectra are shown in Figs.~\ref{fig:lifetime01}--\ref{fig:lifetime15} with simulated line shapes superimposed. 
The fit range has been determined on a case-by-case basis, with the aim of maximizing the coverage while still allowing a simple parametrization of the background as a constant or linear function plus a smoothed step function. 
In some cases, only upper limits on the lifetime could be obtained. %
%
%
The error bars represent our best estimate of the overall experimental uncertainty, including both the statistical uncertainty from the fit and systematic uncertainties. In most cases, the statistical uncertainty dominates. %
The systematic uncertainties were estimated in an ad-hoc fashion by varying the width of the $\alpha$-particle gate, adopting different background parameterizations, widening/narrowing the fit range, and varying selected input parameters for the Monte Carlo simulation, such as the $\gamma$-$\alpha$ angular-correlation function. 
The quality of the experimental data did not warrant a comprehensive and detailed sensitivity-analysis similar to the one reported in Ref.~\cite{galinski2014}.
We note that the non-observation of the $7787\,\textrm{keV}\rightarrow 2051\,\textrm{keV}$ transition in our experiment is consistent with its smaller branching ratio relative to the $7787\,\textrm{keV}\rightarrow 451\,\textrm{keV}$ transition and the higher $\gamma$-ray background at lower energy.
\begin{table*}
\caption{\label{tb:results} Mean lifetimes, $\tau$, and $\gamma$-ray energies, $E_{\gamma}$, determined from line-shape fitting. States are labeled by their excitation energy in keV, as given in the most recent evaluation~\cite{nuclear_data_sheets_A23}. Upper limits are given at the 95\% C.L.}
\begin{ruledtabular}
\begin{tabular}{cclrlc}
\multirow{2}{*}{Initial state}  &  \multirow{2}{*}{Final state(s)}  &  \multirow{2}{*}{$E_{\gamma}$ (keV)}  &  \multicolumn{3}{c}{$\tau$ (fs)} \\\cline{4-6}
 &  &  & \multicolumn{2}{c}{Present} & Literature \\
\hline

2052  &  \phantom{1}451  &  $1601.4\pm 1.3$  &  \multicolumn{2}{c}{$104\pm 18$}  &  $\hphantom{1}80\pm 20$~\cite{engmann71} \\

2771  &  \phantom{1}\textrm{g.s.}   &  $2769.7\pm 1.2$   &  \multicolumn{2}{c}{$\hphantom{0}98\pm 15$}   & $155\pm30$~\cite{engmann71} \\


2908  &  \phantom{1}\textrm{g.s.}   &  $2906.4\pm 1.1$  &  \multicolumn{2}{c}{$\hphantom{0}15\pm 3\hphantom{0}$} &  $<25$~\cite{engmann71} \\

3798  &  \phantom{1}451  &   $3343.9\pm 1.3$   &  \multicolumn{2}{c}{$\hphantom{0}41\pm 6\hphantom{0}$}   &  $<20\times 10^6$~\cite{nuclear_data_sheets_A23}  \\


4353  &  \phantom{1}\textrm{g.s.}   &   $4356\pm 2$   &  \multicolumn{2}{c}{$<11$}    & $<20\times 10^6$~\cite{nuclear_data_sheets_A23} \\



5287  &  \phantom{1}451  &   $4837\pm 3$  &  \multicolumn{2}{c}{$<14$}  &  $\hphantom{00}5\pm2$~\cite{jenkins13}$\hphantom{0}$ \\

5453  &  2052            &   $3401\pm 2$   &  \multicolumn{2}{c}{$<15$}    & \\

6236  &  2052            &   $4186\pm 3$   &  \multicolumn{2}{c}{$<40$}        &  \\

6375  &  2052            &   $4322\pm 6$   &  \multicolumn{2}{c}{$<45$}   &  \\

6899  &  \phantom{1}\textrm{g.s.}   &   $6907\pm 3$   &  \multicolumn{2}{c}{$<10$}        &  \\


7444  &  \phantom{1}\textrm{g.s.}   &   $7443\pm 3$   &  \multicolumn{2}{c}{$<14$}        &  \\

7493  &  2052            &   $5442\pm 2$   &  \multicolumn{2}{c}{$<20$}        &  \\

7787  &  \phantom{1}451             &   $7335\pm 2$   &  \multicolumn{2}{c}{$<12$}        & $\hphantom{1}10\pm3$~\cite{jenkins04}$\hphantom{1}$ \\

\end{tabular}
\end{ruledtabular}
\end{table*}

While we have succeeded in determining several, hitherto unknown, lifetimes in $^{23}$Mg, and placed relevant upper limits on even more, our result for the 7787~keV state, $\tau < 12$~fs at the 95\% C.L., is not sufficiently precise to settle the controversy surrounding the $^{22}$Na$(p,\gamma)$ rate. We note that our result is consistent with the mean lifetime of $\tau=10\pm 3$~fs reported in Ref.~\cite{jenkins04}, but also the mean lifetime of $\tau=2.4^{+0.2}_{-0.3}$~fs implied by the new resonance-strength determination of the Seattle group, $\omega\gamma = 5.7^{+1.6}_{-0.9}$~meV~\cite{sallaska10, sallaska11}, assuming $B_p=0.037\pm 0.07$~\cite{saastamoinen11}.

Our sensitivity has been limited mainly by the low statistics collected on the $7787\,\textrm{keV}\rightarrow 451\,\textrm{keV}$ transition, which may be ascribed to a rather small cross section. Previous to our experiment the cross section for populating the 7787~keV state via ($^3$He,$\alpha$) was unknown. Based on our data we estimate the cross section to be $ \textrm{d}\sigma / \textrm{d}\Omega_{\,\textrm{c.m.}} \approx 3$--4~$\mu$b/sr in the angular range covered in the present experiment ($\theta_{\textrm{c.m.}}>159^{\circ}$) assuming that the $\gamma$ rays are emitted isotropically. %
Our sensitivity has been further limited by the high-energy $\gamma$-ray background resulting from reactions between the beam and the residual gas and water vapor that condensed on the surface of the cooled target, cf.~Section~\ref{sec:setup}.

\begin{figure}
\includegraphics[width=0.95\linewidth, angle=0, clip=true, trim= 10 20 60 70]{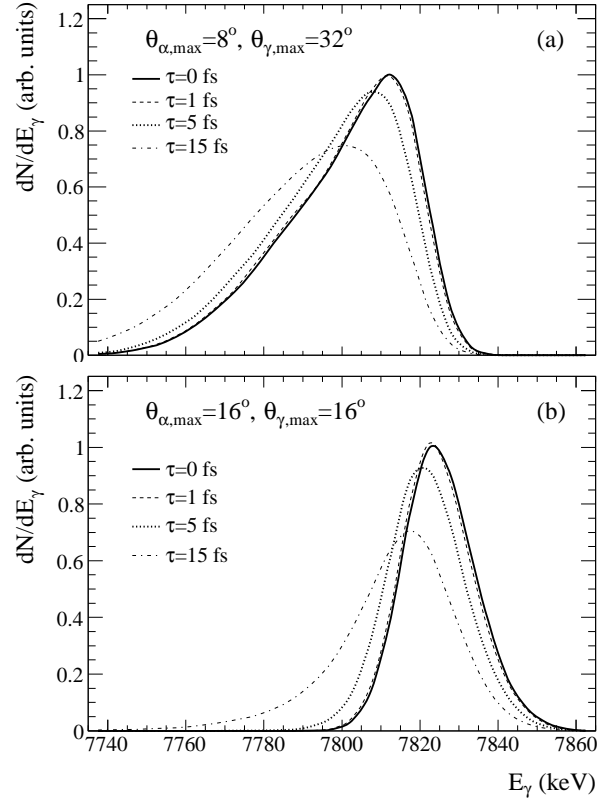}%
\caption{\label{fig:mcanal}(a) Simulated line shape of the $7787\,\textrm{keV}\rightarrow 451\,\textrm{keV}$ transition with the current experimental setup assuming mean lifetimes of 0, 1, 5, and 15~fs. (b) Same, but with the acceptance of the $\Delta E$-$E$ telescope doubled and the acceptance of the HPGe detector halved.}
\end{figure}

It is also worth considering if the geometry of the detector setup can be improved. %
Fig.~\ref{fig:mcanal}~(a) shows the simulated line shape of the $7787\,\textrm{keV}\rightarrow 451\,\textrm{keV}$ transition with the current experimental setup assuming mean lifetimes of 0, 1, 5, and 15~fs. In Fig.~\ref{fig:mcanal}~(b) the same four lifetimes have been plotted, but with a modified setup: the angular acceptance of the $\Delta E$-$E$ telescope has been doubled while the angular acceptance of the HPGe detector has been halved (thus keeping the product $\Omega_{\alpha}$$\Omega_{\gamma}$ nearly constant). 
Based on the simulated line shapes it would appear that the lifetime sensitivity can be improved by simply reducing the acceptance of the HPGe detector, while increasing the acceptance of the $\Delta E$-$E$ telescope to maintain the same level of statistics. It must be remembered, however, that the increased acceptance of the $\Delta E$-$E$ telescope will result in a worse excitation-energy resolution. The gates in the $\alpha$-particle spectrum will have to be made wider and consequently more background will appear in the $\alpha$-gated $\gamma$-ray spectra. Also, the increased acceptance of the $\Delta E$-$E$ telescope will increase the sensitivity to the angular distribution of the $\alpha$-particles, which is usually not known. 
Thus, the preferred solution would be to keep the acceptance of the $\Delta E$-$E$ telescope fixed and employ several HPGe detectors, rather than just one, so that the HPGe detectors can be placed farther away, or employ a position sensitive HPGe detector such as the one described in Ref.~\cite{schumaker07}. This would, however, complicate both the experiment and the data analysis.

\section{Conclusion}

The $^{22}$Na$(p,\gamma)$ reaction plays an important role in classical novae by limiting the production of the long-lived $\gamma$-ray emitter $^{22}$Na~\cite{jose06}. At the peak temperatures occurring in classical novae, the reaction rate is dominated by a single resonance due to an excited state in the compound nucleus, $^{23}$Mg, at an excitation energy of $7787$~keV. The two direct measurements of the resonance strength reported in the literature differ by a factor of 3~\cite{stegmuller96, sallaska10}. The resonance strength deduced indirectly from the properties of the $7787$~keV state is consistent with the lower of the two reported direct measurements, but hinges on a single lifetime measurement with a 30\% error bar~\cite{jenkins04}. %
Here we have reported on a new lifetime measurement performed at TRIUMF using the DSAM technique~\cite{branford73, alexander78}. %
We have developed a Monte Carlo simulation program to model the measured $\gamma$-ray line shapes. Our program improves on a similar program previously used at TRIUMF~\cite{galinski2014} by including the effects of multiple scattering.
We have successfully determined several lifetimes in $^{23}$Mg for the first time. Our result for the 7787~keV state, $\tau < 12$~fs at the 95\% C.L., is unfortunately not sufficiently precise to settle the controversy surrounding the $^{22}$Na$(p,\gamma)$ rate. 
When combined with the proton-decay branching-ratio determination of Ref.~\cite{saastamoinen11}, $B_p = 0.037\pm 0.007$, and assuming a spin of $J=7/2$, our upper limit on the lifetime of the 7787~keV state gives a lower limit on the resonance strength of $\omega\gamma > 1.1$~meV at the 95\% C.L.

\begin{acknowledgments}
We thank W.~N.~Lennard for helpful discussions on the properties of the $^3$He-implanted Au target, and we thank E.~Rand for supplying us with the GEANT4 simulation of the HPGe detector.
OSK acknowledges support from the Villum Foundation.
TRIUMF receives federal funding via a contribution agreement through
the National Research Council of Canada.
\end{acknowledgments}

\bibliography{okirsebom}

\end{document}